\documentclass[twocolumn,twoside,slac]{revtex4}
\usepackage{graphicx}
\usepackage{fancyhdr}
\usepackage[latin1]{inputenc}
\begin{document}

\title{FPGA Co-processor for the ALICE High Level Trigger}

\author{G.~Grastveit$^1$, H. Helstrup$^2$, V. Lindenstruth$^3$, C. Loizides$^4$, D. Roehrich$^1$, B. Skaali$^5$, \\ T. Steinbeck$^3$, R. Stock$^4$, H. Tilsner$^3$, K. Ullaland$^1$, A. Vestbo$^1$ and T. Vik$^5$ \\ for the ALICE Collaboration \\ \vspace{0.5cm}}

\affiliation{$^1$ Department of Physics, University of Bergen, Allegaten 55, N-5007 Bergen, Norway}
\affiliation{$^2$ Bergen University College, Postbox 7030, N-5020 Bergen, Norway}
\affiliation{$^3$ Kirchhoff Institut für Physik, Im Neuenheimer Feld 227, D-69120 Heidelberg, Germany}
\affiliation{$^4$ Institut für Kernphysik Frankfurt, August-Euler-Str. 6, D-60486 Frankfurt am Main, Germany}
\affiliation{$^5$ Department of Physics, University of Oslo, P.O.Box 1048 Blindern, N-0316 Oslo, Norway}

\begin{abstract}
The High Level Trigger (HLT) of the ALICE experiment requires massive parallel computing. One of the main tasks of the HLT system is two-dimensional cluster finding on raw data of the Time Projection Chamber (TPC), which is the main data source of ALICE. To reduce the number of computing nodes needed in the HLT farm, FPGAs, which are an intrinsic part of the system, will be utilized for this task. VHDL code implementing the \emph{fast cluster finder} algorithm, has been written, a testbed for functional verification of the code has been developed, and the code has been synthesized.
\end{abstract}

\maketitle

\thispagestyle{fancy}

\section{Introduction}

The central detectors of the ALICE experiment \cite{alice}, mainly its large Time Projection Chamber (TPC) \cite{alicetpc}, will produce a data size of up to 75 MByte/event at an event rate of up to 200 Hz resulting in a data rate of $\sim$15 GByte/sec. This exceeds the foreseen mass storage bandwidth of 1.25 GByte/sec by one order of magnitude. 
The High Level Trigger (HLT), a massive parallel computing system, processes the data online doing pattern recognition and simple event reconstruction almost in real-time, in order to select interesting (sub)events, or to compress data efficiently by modeling techniques \cite{rt2003}, \cite{berger02}. 
The system will consist of a farm of clustered SMP-nodes based on off-the-shelf PCs connected with a high bandwidth low latency network. The system nodes will be interfaced to the front-end electronics via optical fibers connecting to their internal PCI-bus, using a custom PCI receiver card for the detector readout (RORC) \cite{anders}.

Such PCI boards, carrying an ALTERA APEX FPGA, SRAM, DRAM and a CPLD and FLASH for configuration have been produced and are operational. Most of the local pattern recognition is done using the FPGA co-processor while the data is being transferred to the memory of the corresponding nodes. 






Focusing on TPC tracking, in the conventional way of event reconstruction one first calculates the cluster centroids with a \emph{Cluster Finder} and then uses a \emph{Track Follower} on these space points to extract the track parameters \cite{yepes96}, \cite{startrigger}. Conventional cluster finding reconstructs positions of space points from raw data, which are interpreted as the crossing points between tracks and the center of padrows. The cluster centroids are calculated as the weighted charge mean in pad and time direction. The algorithm typically is suited for low multiplicity data, because overlapping clusters are not properly handled. By splitting clusters into smaller clusters at local minima in time or pad direction a simple deconvolution scheme can be applied and the method becomes usable also for higher multiplicities of up to dN/dy of 3000 \cite{rt2003}. 

To reduce data shipping and communication overhead in the HLT network system, most of the cluster finding  will be done \emph{locally} defined by the readout granularity of the readout chambers. The TPC front-end electronic defines 216 data volumes, which are being read out over single fibers into the RORC. 

We therefore implement the \emph{Cluster Finder} directly on the FPGA co-processor of these receiving nodes while reading out the data. The algorithm is highly local, so that each sector and even each padrow can be processed independently, which is another reason to use a circuit for the parallel computation of the space points.

\section{The Cluster Finder Algorithm}

\begin{figure}
\begin{center}
\includegraphics[width=6cm]{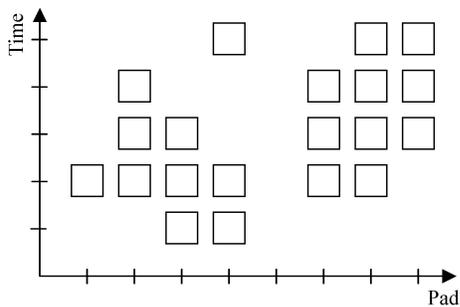}
\end{center}
\caption{Input charge values in the time-pad plane}
\label{algo1b}
\end{figure}

The data input to the \emph{cluster finder} is a zero suppressed, time-ordered list of charge values on successive pads and padrows. The charges are ADC values in the range 0 to 1023 drawn as squares in figure~\ref{algo1b}. By the nature of the data readout and representation,  they are placed at integer time, pad and padrow coordinates. 

Finding the particle crossing points breaks down into two tasks: Determine which ADC values belong together, forming a cluster, and then calculate the center of gravity as the weighted mean using
\begin{eqnarray}
G_P & = & \left. \sum q_i p_i \, \right/ \, Q \label{cogpad} \\
G_T & = & \left. \sum q_i t_i \, \right/ \, Q \label{cogtime} \,,
\end{eqnarray}
where $G_T$ is the center of gravity in the time-direction, $G_P$ in the pad direction, $Q=\sum q_i$ is the total charge of the cluster.

\subsection{Grouping charges into sequences}\label{groupingcharges}
Because of the ALTRO\cite{altro} data format, we choose to order the input data first along the time direction into \emph{sequences}, before we merge sequences on neighboring pads into clusters. 

\begin{figure}
\begin{center}
\includegraphics[width=6cm]{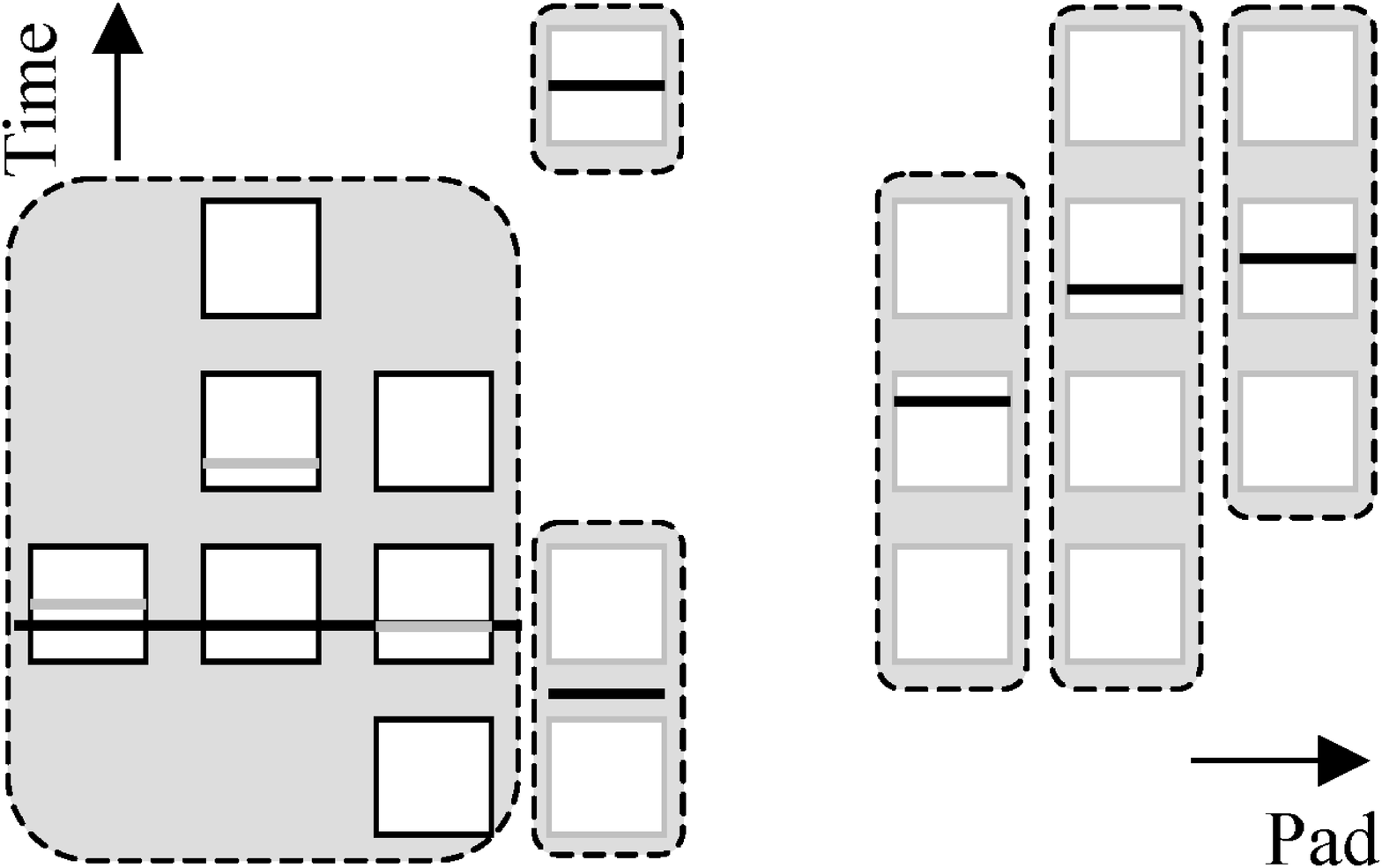}
\end{center}
\caption{Grouping input data into sequences, and merging neighboring sequences into clusters.}
\label{algo2}
\label{algo3}
\end{figure}

A sequence is regarded as a continuous (vertical) stack of integer numbers. They are shown in figure~\ref{algo2}. For each sequence the center of gravity is calculated using eqn.~(\ref{cogtime}). The result is a non-integer number. This is illustrated by the horizontal black lines.

\subsection{Merging neighboring sequences}\label{mergingsequences}
Searching along the pad direction (from left to right in the figures), sequences on adjacent pads are merged into a starting cluster if the difference in the center of gravity is less than a selectable \emph{match distance} parameter\footnote{Typically we set the match distance parameter to 2.}. The center of gravity value of the last appended (rightmost) sequence is kept and used when searching for matches on the following pad, also shown in figure~\ref{algo3}.

\begin{figure}
\begin{center}
\includegraphics[width=7cm]{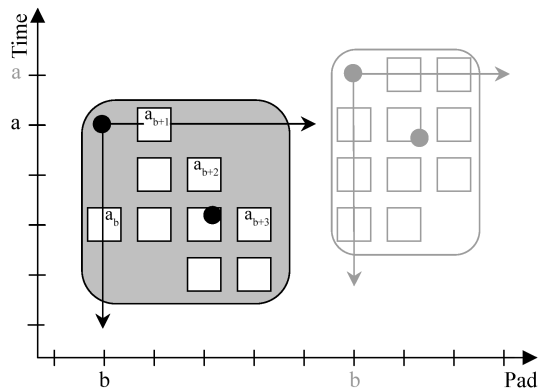}
\end{center}
\caption{Finished clusters in the time-pad plane, where the black dots in the middle mark the centroid. $a$ and $b$ specify the absolute position of the cluster, $a_i$ are the time positions of the sequences that form the cluster.}
\label{relscale}
\end{figure}

A cluster that has no match on the next pad is regarded as finished. If it is above the noise threshold (requirements on size and total charge), the final center of gravities in time and pad according to eqns.~(\ref{cogpad}) and (\ref{cogtime}) are calculated. This is --besides other definitions-- illustrated in figure~\ref{relscale}. The space-point, given by these two values transformed into global coordinate system, and the total charge of the cluster are then used for tracking [6].


\section{Algorithm adaptions for VHDL}

\begin{figure}
\begin{center}
\includegraphics[width=7cm]{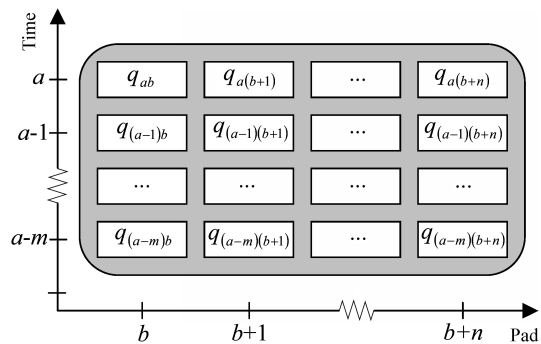}
\end{center}
\caption{The definition of a general cluster used to simplify the algorithm for the FPGA}
\label{gencluster}
\end{figure}

To transform the algorithm into a version which is suitable for a computation in hardware, we 
define a general cluster as a matrix in the time-pad plane according to figure~\ref{gencluster}.

Of course, clusters can be any size and shape and can be placed anywhere within the time-pad plane. The upper left corner is chosen as the reference point since this charge will enter the cluster finder circuit first. The values $a$ and $b$ define the absolute placement of the corner, the size of the cluster is given by $m$ and $n$. Allowing charge values to be zero covers varying shapes, but since only one sequence per pad is merged, zero charge values in the interior of the clusters can not occur.

Using the general cluster definition, the total charge $Q$ is the sum of the total sequence charges $Q_j$ according to
\begin{equation}
Q = \sum_{j=b}^{b+n} Q_j \hspace{0.5cm} \text{where} \hspace{0.5cm} Q_j = \sum_{i=a-m}^{a} q_{ij} \label{totseqcharge}
\end{equation}
and the formulas for the center of gravity change into
\begin{eqnarray}
G_P & = & \left. \sum_{j=b}^{b+n} j Q_j \right/ Q \label{cogpad2}\\
G_T & = & \left. \sum_{j=b}^{b+n}\sum_{i=a-m}^{a} i q_{ij} \right/ Q \label{cogtime2}
\end{eqnarray}

By calculating the center of gravity relative to the upper left corner of the cluster, the multiplicands $i$ and $j$ in (\ref{cogpad2}) and (\ref{cogtime2}), which start counting from $a$ and $b$ respectively, are exchanged by indexes starting at 0. Using the definitions (\ref{totseqcharge}) and figure~\ref{relscale}, we get for the center of gravities the following two equations:
\begin{eqnarray}
G_P & = & b + \left. \sum_{k=0}^{n} k Q_{b+k} \right/ Q \label{cogpad3}\\
G_T & = & a - \left. \sum_{j=b}^{b+n}\sum_{k=0}^{m} k q_{(a_j-k)j} + (a-a_j) Q_j \right/ Q\label{cogtime3} 
\end{eqnarray}

These formulas are better suited for the FPGA implementation because the range of the multiplicands are restricted for all multiplications. $(a-a_j)$ and $k$ are within the height and width of a cluster. 

When the relative centroid has been determined, it has to be transformed to the absolute coordinate system (using $a$ and $b$). Thus,  the FPGA implementation has to calculate $a$, $b$, $Q$ and the result of the two sums. Per cluster these 5 integer values will be sent to the host, which in turn uses (\ref{cogpad3}) and (\ref{cogtime3}) for the final result.

\section{VHDL implementation}

\begin{figure}
\begin{center}
\includegraphics[width=7cm]{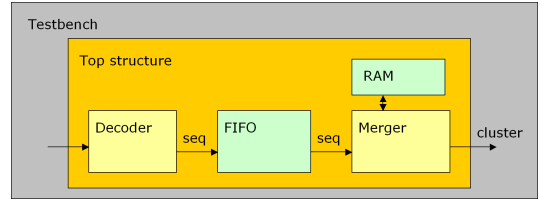}
\end{center}
\caption{The block diagram of the FPGA cluster finder}
\label{blockdiagram}
\end{figure}

The FPGA implementation has four components as shown in figure~\ref{blockdiagram}. The two main parts are the \emph{Decoder}, which groups the charges into sequences (section \ref{groupingcharges}), and the \emph{Merger}, which merges the sequences into clusters (section \ref{mergingsequences}).
 
\subsection{Decoder}
The \emph{Decoder} handles the format of the incoming data and computes properties of the sequences. We see from (\ref{cogtime3}) that two parts are ``local'' to every sequence. These are $Q_j$ and $\sum_{k=0}^{m} k q_{(a_j-k)j}$, where the total charge of the sequence is also used in (\ref{cogpad3}). Calculation of these two sequence properties need the individual charges $q_{ij}$ but does not involve computations that require several clock cycles. The two properties are therefore computed as the charges enter the circuit, so that the charges have not to be stored. Additionally, as required by the algorithm, the center of gravity of the sequences needs to be calculated. Only calculating the geometric middle, is sufficiently precise and faster than the exact computation.

When all information about a sequence is assembled, it will be sent to the second part, the \emph{Merger}. That way the narrow, fixed-rate data stream is converted into wide data words of varying rate. That rate is dependent on the length of the sequences. Long sequences give low data rate and vice versa. Because of this, and also because of varying processing speed of the \emph{Merger}, a small FIFO is used to buffer the sequence data. Since the \emph{Decoder} handles data in real time as they come, the sequences will also be ordered by ascending row, ascending pad and descending time.

\subsection{Merger}
The \emph{Merger} decides which sequences belong together and merges them. The merging in effect increases the $k$ index of (\ref{cogpad3}) and $j$ index of (\ref{cogtime3}) by one, adding one more sequence to the sums in the numerators. It also updates the total charge of the cluster $Q$. 

\begin{figure}
\begin{center}
\includegraphics[width=4cm]{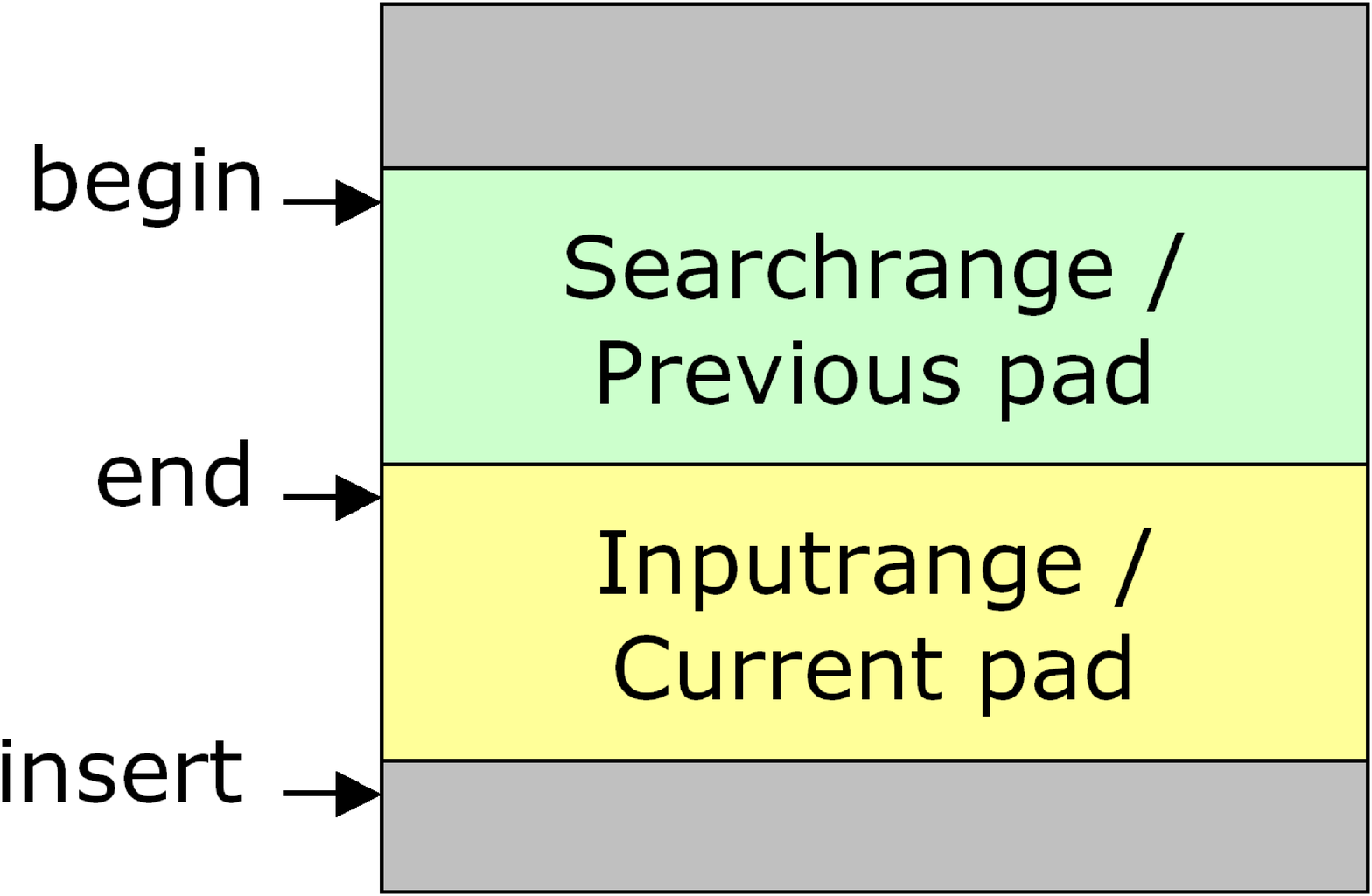}
\end{center}
\caption{The ring buffer storing clusters of the previous and the actual pad.}
\label{ringbuffer}
\end{figure}

Since the \emph{Merger} is processing the data in arriving order, merging will be done with sequences on the preceding pads. And more precisely, since by definition a cluster does not have holes, only the immediately preceding pad needs to be searched. Therefore we need only two lists in memory (see figure~\ref{ringbuffer}). One list contains the started cluster of the previous pad; the other list contains started clusters already processed at the current pad. Clusters are removed from the search range when a match is found or when the cluster is finished. Clusters are inserted in the input range after merging or when starting a new cluster. The list of clusters on the current pad, the input range, will be searched when a sequence on the next pad arrives. When that happens, there will be no matches for the remaining clusters on the preceding pad. Hence we output clusters in the old search range and exchange the lists. The old list becomes free memory, the current list becomes the old, and a new current list with no entries is created. At the end of a row or when a pad is skipped the clusters in both the lists must be finished. All the clusters are sent and both the lists are emptied. The two lists are implemented as a ring buffer stored in a dual-port RAM. The beginning and ending  of the lists is marked by three memory pointers (\emph{begin}, \emph{end} and \emph{insert} pointer). 

\begin{figure}
\begin{center}
\includegraphics[width=7cm]{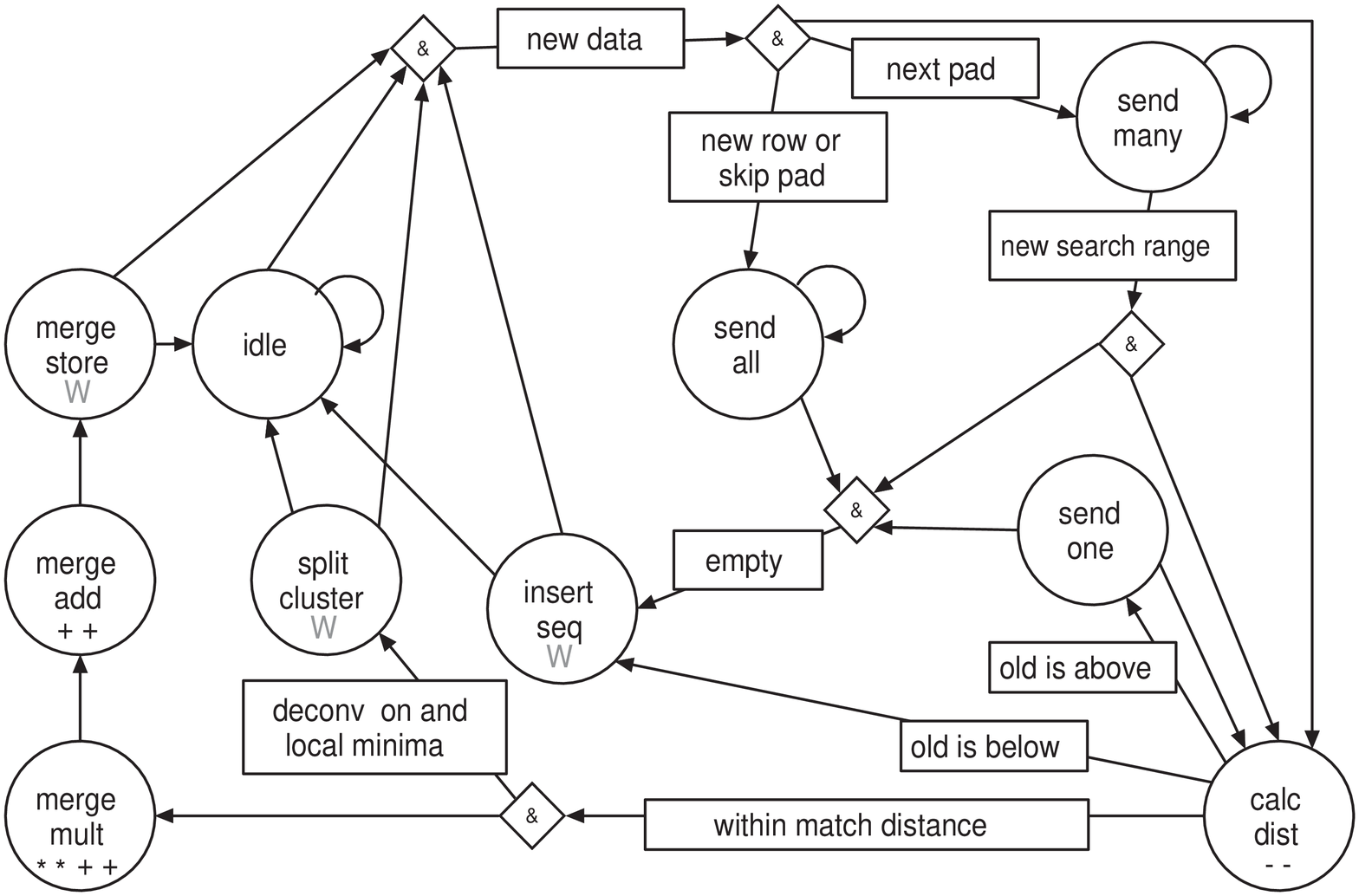}
\end{center}
\caption{The state machine of the \emph{Merger}. The undertaken arithmetical operations and write accesses are marked.}
\label{mergersm}
\end{figure}

The state machine of the \emph{Merger} is shown in figure~\ref{mergersm}. To prevent overflowing the FIFO, the number of states the \emph{Merger} machine needs to visit for each sequence is kept as low as possible. At the same time every state must be as simple as possible to enable a high clock rate. The ring buffer causes systematic memory accesses; read and write addresses are either unchanged or incremented. To keep the number of states down, the computations of merging are done in parallel, thereby using more of the FPGA resources. The resources needed are two adders (signed), and two ``smart multipliers'' (see \ref{smartmult}). 

There are three different types of states: States that remove a cluster from the search range sending it to the output (or discarding it if it is a noise cluster or overflow occurred), states that do arithmetic and states that insert a new cluster into the list of the current pad. 

Arithmetic is done in three cases. When a new sequence on the current row and pad enters, the distance between starting times, $(a - a_j)$, is calculated. Because the data type is unsigned, $a$ is kept at the top of the cluster by assigning it the highest of the starting times. The roles of the unfinished cluster and sequence are interchangeable. The distance between the middle of the incoming and the middle of the first in the search range is also calculated (\emph{calc\_dist}). If the result is within match distance, merging occurs.

In the \emph{merge\_mult} state the last part in (\ref{cogtime3}), is computed for the lower of the unfinished cluster and the incoming sequence. The rest of the calculations are done unconditionally. They are finished in the \emph{merge\_add} state. After merging the resulting cluster is inserted into the current list (\emph{merge\_store}), at the same time the old unfinished cluster is removed from the search range by incrementing the \emph{begin} pointer.

If a match is not found in the \emph{calc\_dist} state the ordering of data is crucial: The first cluster in the search range is always the one of highest time value, so if it is below the incoming, the rest of the clusters in the search range will also be below. Hence there can be no matches for the incoming sequence and it can be inserted into the current list (\emph{insert\_seq}).
In the opposite case, if the cluster in the search range is higher than the incoming, all subsequent incoming sequences will be below or on other pads, so the cluster in search range must be finished. The cluster is output, and the \emph{begin} pointer is incremented (\emph{send\_one}).
The same procedure happens in three other states. By definition, on a change of row there will be no more matches neither in the search range nor the current list. Therefore all the clusters are sent (\emph{send\_all}). As described above, motivating the ring buffer: when there is a change of the pad, clusters in the search range are sent and the lists are renamed (\emph{send\_many}). The last case finishing a cluster occurs if convolution is turned on and a local minima in a cluster is detected. The \emph{split\_cluster} state is combination of \emph{send\_one} and \emph{insert\_seq}. The old cluster is sent to the output and the incoming is inserted into the search range.

\subsection{SmartMult}\label{smartmult}
Both the \emph{Decoder} and the \emph{Merger} do multiplications where the range of the multiplicand is limited to the size of the cluster as intended by (\ref{cogpad3}) and (\ref{cogtime3}). To save resources (logic cells) on the FPGA and increase the clock speed, we replace the standard multipliers by \emph{SmartMult}, which takes the multiplicand and multiplier as arguments and uses left shifts and one or two adders for the calculations. The multiplicand is limited to reduce the number of shifts needed. 
The limit of the multiplicand determines the highest sequence that is allowed and the highest number of merges possible. Clusters larger than this are flagged as overflowed and ignored when finished. 



\begin{figure}
\begin{center}
\includegraphics[width=8cm]{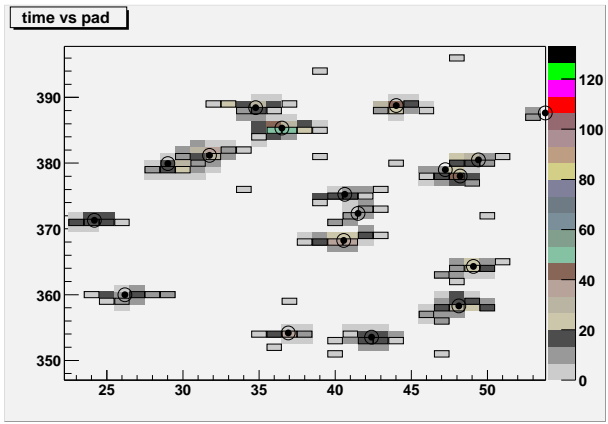}
\end{center}
\caption{A part of a time-pad plane. Patches are input charges, squares mark the geometric middle of the sequences, points and circles mark the centroids found by the FPGA and the C code resp.}
\label{verification}
\end{figure}

\begin{table}
\caption{Distribution of clock cycles spent in the various states of the \emph{Merger},
corresponding to dN/dy of 2500 and 1000 and with/without deconvolution}
\label{statetime}
\begin{center}
\begin{tabular}{|c|c|c|c|c|}
\hline 
{\bf state} & {\bf 2500-n} & {\bf 2500-de} & {\bf 1000-n} & {\bf 1000-de} \\
\hline
\hline
  idle          &  26,0 &   24,9  &  31,0 &  30,0 \\
merge\_mult     &  6,5  &   6,6  &  9,1  &   9,0 \\
merge\_add      &  6,5  &   6,6 &  9,1  &   9,0 \\
merge\_store    &  6,5  &   6,6  &  9,1  &   9,0 \\
send\_all       &  0,5  &   0,5   &  0,5  &   0,5 \\ 
send\_many      &  5,4  &   5,4   &  5,6  &   5,6 \\
send\_one       &  9,3  &   9,3   &  5,6  &   5,7 \\
calc\_dist      &  26,9 &   27,2  &  21,9 &   22,4 \\
insert\_seq     &  12,6 &   12,7   &  8,2  &   8,5 \\
split\_cluster  &   0,0 &   0,0   &  0,0  &   0,2 \\
\hline
\end {tabular}
\end{center}
\end{table}

\enlargethispage{\baselineskip}

\subsection{Verification}
For verification purposes the system is stimulated by a testbench. The testbench reads an ASCII file containing simulated ALIROOT raw data in an ALTRO like back-linked list, which is sent to the \emph{Decoder}. Operation of the circuit is then studied in a simulator for digital circuits. Found clusters of the \emph{Merger} circuit are directed back to the testbench which writes the results to a file. That file is then compared to a result file made by a C++ program running nearly the HLT C++ algorithm on the same simulated input data. Since the number of clusters is in the order of thousands and to eliminate human error, another C++ program compares the two result files. The found clusters agree as is graphically shown in figure~\ref{verification}.

\subsection{Timing}
The circuit has been synthesized and currently uses 1937 logic cells. That is 12\% of the available resources on the APEX20KE-400-2x, which we are using for prototyping. The clock speed is 35 MHz. For the different input data sets taken for the verification, the \emph{Merger} has been in the \emph{idle} state more than 24\% of the time. For two different data sets the distribution of the clock cycles is shown in table~\ref{statetime} for dN/dy of 2500 and 1000 and with/without deconvolution. As merging is the time critical part of the circuit, there is a safety margin for higher multiplicity data. 

\section{Conclusion}

We have developed a fast cluster finding algorithm in VHDL, suitable for implementation in an FPGA. On an APEX20KE-400-2x FPGA from ALTERA, the synthesized circuit uses 12\% (1937) of the logic cells and runs at a clock speed of 35 MHz. We measured the time critical paths (in the \emph{Merger} circuit) to be idle more than 24\% of the time.


\end{document}